\documentclass[12pt]{article}
\usepackage[]{epsfig}
\usepackage[centertags]{amsmath}
\usepackage{amsfonts}
\usepackage{amssymb}
\usepackage[english]{babel}

\begin{document}

\title{\bf On the energy crisis in noncommutative $CP(1)$ model }
\author{
Lucas Sourrouille$^a$
\\
{\normalsize \it $^a$Departamento de F\'\i sica, FCEyN, Universidad
de Buenos Aires}\\ {\normalsize\it Pab.1, Ciudad Universitaria,
1428, Ciudad de Buenos Aires, Argentina}
\\
{\footnotesize  sourrou@df.uba.ar} } \maketitle

\abstract{We study the $CP(1)$ system in (2+1)-dimensional noncommutative space with and without Chern-Simons term. Using the Seiberg-Witten map we convert the noncommutative $CP(1)$ system to an action written in terms of the commutative fields. We find that this system presents the same infinite size instanton solution as the commutative Chern-Simons-$CP(1)$ model without a potential term. Based on this result we argue that the BPS equations are compatible with the full variational equations of motion, rejecting the hypothesis of an "energy crisis". In addition we examine the noncommutative $CP(1)$ system with a Chern-Simons interaction. In this case we find that when the theory is transformed by the Seiberg-Witten map it also presents the same instanton solution as the  commutative Chern-Simons-$CP(1)$ model. }


\newpage


\vspace{1cm}
\section{Introduction}

The study of field theories in noncommutative space has received much attention in the last few years \cite{rew}. The connection between these theories and string theory was first considered by Connes, Douglas and Schwartz, who observed that noncommutative geometry arise as a possible scenario for certain low energy of string theory and M-theory. Afterwards Seiberg and Witten showed that the low energy dynamics of string theory can be described in terms of the noncommutative Yang-Mills theory \cite{witten1}. Since then many papers appeared covering diverse applications of noncommutative theory in physical problems. One important aspect in these investigations has been the study of noncommutative instantons and solitons. These play a fundamental role in understanding of the nonperturbative effects in noncommutative gauge theory. In four-dimensional Yang-Mills theory it was shown the existence of instantons even in the $U(1)$ case \cite{nek}, and nontrivial solitons were found in noncommutative scalar theory \cite{strom}.

The two-dimensional $CP(N)$ sigma model presents many similarities to four-dimensional Yang-Mills theory.  An important one is the existence of the instanton solutions \cite{witten}. The extension of this similarity to the case of noncommutative space was first analyzed by B. Lee, K. Lee and S. Yang in Ref.\cite{yang}. After that many interesting properties of the $CP(1)$ model in noncommutative space were found \cite{varios}. The noncommutative $CP(1)$ model was investigated by S. Ghosh using the Seiberg-Witten map in Ref.\cite{ghosh}. The author utilizes there the Seiberg-Witten map to convert the noncommutative $CP(1)$ action to an action written in terms of the commutative fields. He claims that in the subsequent theory the BPS equations are not compatible with the variational equation of motion, attributing this particular feature to an inadequate definition of energy-momentum tensor in noncommutative space. In this paper we show, using the results recently obtained in Ref.\cite{my}, that the BPS and Euler-Lagrange equations obtained in Ref.\cite{ghosh} are the same as the well-know BPS and motion equations of the commutative $CP(1)$ theory. Using the method explored in Ref.\cite{ghosh} we also study the case of noncommutative Chern-Simons-$CP(1)$  theory without a potential term. We find that both models share the same instanton solution, which is the infinite size solution found in Ref.\cite{my} for commutative Chern-Simons-$CP(1)$ in the  absence of potential term.

The paper is organized as follows:
In Section $2$ we review the result obtained in Ref.\cite{my}. Section $3$  is advocated to show that the BPS and variational motion equation present in the model study by Ghosh are compatible. Finally we analyze in Section $4$ the Chern-Simons-$CP(1)$ case.

\section{The commutative Chern-Simons-$CP(1)$ instanton}

In this Section we shall describe briefly the characteristics of the Chern-Simons-$CP(1)$ model in commutative space, which was studies in Ref.\cite{my}.
We begin by considering a $(2+1)$-dimensional Chern-Simons model coupled to a complex two component field $n(x)$  described by the action

\begin{eqnarray}
S&=& S_{cs}+\int d^3 x |D_\mu n|^2
\label{S1}
\end{eqnarray}

Here  $D_{\mu}= \partial_{\mu} - iA_{\mu}$ $(\mu =0,1,2)$ is the covariant derivative and $S_{cs}$ is the Chern-Simons action given by

\begin{eqnarray}
 S_{cs}= \kappa\int d^3 x \epsilon^{\mu \nu \rho} A_\mu \partial_\nu A_\rho= 2\kappa \int d^3 x \left( A_0 F_{12} + A_2 \partial_0 A_1
\right)
\end{eqnarray}

where
\begin{eqnarray}
F_{\mu \nu}=\partial_{\mu}A_{\nu}-
\partial_{\nu}A_{\mu} \label{F}
\end{eqnarray}

The metric signature is $(1,-1,-1)$ and  the two component field $n(x)$ is subject to
the constraint $n^\dagger n = 1$. The constraint can be introduced in the variational process with a Lagrange multiplier. Then we extremise the following action

\begin{eqnarray}
S&=& S_{cs}+\int d^3 x |D_\mu n|^2 + \lambda (n^\dagger n -1)
\label{}
\end{eqnarray}

The obtained field equations are

\begin{eqnarray}
D_\mu D^\mu n +\lambda n =0
\label{motion1}
\end{eqnarray}
and

\begin{eqnarray}
A_\mu = \frac{1}{2} \left(-2i n^\dagger \partial_\mu n - \kappa\epsilon_{\mu \nu \rho} F^{ \nu \rho}\right)
\label{motion1}
\end{eqnarray}

From the first equation we can deduce

\begin{eqnarray}
\lambda= (n^\dagger D_\mu D^\mu n)n
\end{eqnarray}

to yield
\begin{eqnarray}
D_\mu D^\mu n = (n^\dagger D_\mu D^\mu n)n
\label{motion2}
\end{eqnarray}

The equation of notion for the gauge field can be rewritten as  $\kappa\epsilon_{\mu \nu \rho} F^{ \nu \rho} =  J_\mu$ where $J_\mu = -i [n^\dagger D_\mu n - n(D_\mu n)^\dagger]=2\Big(-in^\dagger \partial_\mu n - A_\mu \Big)$ is the matter current. The time component of Eq.(\ref{motion1}),

\begin{eqnarray}
2\kappa F_{12} =  -J_0 \label{gauss}
\end{eqnarray}

is Gauss's law of the Chern-Simons dynamics\cite{jackiw}.

Using the Gauss's law, the energy functional for a static field configuration can be expressed as

\begin{eqnarray}
E=  \int d^2 x \Big(\kappa^2 B^2 + |D_i n|^2  \Big) \,,
\;\;\;\;\;\
i = 1,2  \label{statich}
\end{eqnarray}
where $B=F_{1 2}$.
Using the identity $|D_i n|^2 = |( D_1 \pm iD_2)n|^2 \mp B \pm \frac{\epsilon^{ij}}{2} \partial_i J_j$ , the energy (\ref{statich}) becomes

\begin{eqnarray}
E=\int d^2 x \,\, \Big( \kappa^2 B^2 +  |( D_1 \pm iD_2)n|^2  \Big) \mp 2\pi Q_{CP(1)}
\label{H2}
\end{eqnarray}
where $Q_{CP(1)}$ is the topological bound of the pure CP(1) system \cite{witten},\cite{my}. The saturation of the bound take place when the first-order Bogomolnyi self-duality equations are satisfied,

\begin{eqnarray}
|( D_1 \pm iD_2)n|^2 =0
\,,
\;\;\;\;\;\
B=0
\label{bogo}
\end{eqnarray}

It was shown in Ref.\cite{my} that the N-solution for these equations are instantons of infinite size. The infinite size of the solution is due to the condition $B=0$ derived from the introduction of the Chern-Simons term in the $CP(1)$ model.

\section{The noncommutative $CP(1)$ model and the Seiberg-Witten map  }

We shall consider the Chern-Simons-$CP(1)$ model before introduced, but now on (2+1)-dimensional noncommutative space. The coordinates of the noncommutative space obey the relations

\begin{eqnarray}
\left[x_1, x_2 \right]= i\theta
\,,
\;\;\;\;\;\
\left[x_i, t \right]= 0
\label{NC}
\end{eqnarray}

with $\theta >0$. It will be convenient introduce complex variables $z$ and $\bar{z}$

\begin{eqnarray}
z=\frac{1}{\sqrt{2}}\Big(x^1 +ix^2\Big)
\,,
\;\;\;\;\;\
\bar{z}=\frac{1}{\sqrt{2}}\Big(x^1 -ix^2\Big)
\end{eqnarray}

which can be related to an annihilation and creation operators $\hat{a}$ and $\hat{a}^\dagger$ acting on a Fock space,

\begin{eqnarray}
\hat{a}=\frac{1}{\sqrt{\theta}}z
\,,
\;\;\;\;\;\
\hat{a}^\dagger=\frac{1}{\sqrt{2}}\Big(x^1 -ix^2\Big)
\end{eqnarray}

so that (\ref{NC}) becomes

\begin{eqnarray}
\left[\hat{a},\hat{a}^\dagger \right]= 1
\end{eqnarray}

In this way, through the action of $\hat{a}^\dagger$ on the vacuum state $|0\rangle$, eigenstates of the number operator

\begin{eqnarray}
\hat N = a^\dagger a
\label{number}
\end{eqnarray}

are generated.
With our conventions, derivatives in the Fock space are given by

\begin{eqnarray}
\partial_z = -\frac{1}{\sqrt \theta}
[\hat a^\dagger,~] \; , \;\;\;\;\;\; \partial_{\bar z} =
\frac{1}{\sqrt \theta} [\hat a,~] \;, \label{5} \end{eqnarray}

and integration on the noncommutative plane should be interpreted
as a trace

\begin{eqnarray} \int d^2x  \to 2\pi \theta {\rm Tr}  \;. \end{eqnarray}

The action of the noncommutative $CP(1)$ model \cite{ghosh} is

\begin{eqnarray}
\hat{S}&=& \int d^3 x  |\hat{D}_\mu \hat{n}|^2
\label{NCP1}
\end{eqnarray}
where the covariant derivative is defined as

\begin{eqnarray}
\hat{D}_{\mu}\hat{n}= \partial_{\mu}\hat{n} - ie\hat{A}_{\mu}\hat{n}
\end{eqnarray}

and with $\hat{n}$ and $\hat{A}_{\mu}$ we indicate that the fields are  noncommutative.
The new gauge fields transforms as follows
\begin{eqnarray}
\hat{\delta} \hat{n} =i\hat{\alpha} \hat{n}
\,,
\;\;\;\;\;\
\hat{\delta} \hat{A_\mu} =\partial_\mu \hat{\alpha} +i\Big[\hat{\alpha}, \hat{A_\mu}\Big]
\end{eqnarray}

The noncommutative fields $\hat{n}$, $\hat{A}_{\mu}$ and the noncommutative gauge parameter $\hat{\alpha}$, can be expressed as a function of the ordinary fields $n$, $A_\mu$ and the commutative gauge parameter $\alpha$ by the so called Seiberg-Witten \cite{witten1} map, whose expression at the lowest order in $\theta$ is

\begin{eqnarray}
\hat{A_\mu} = A_\mu + \theta^{\sigma \rho} A_\rho \Big( \partial_\sigma A_\mu - \frac{1}{2}\partial_\mu A_\sigma \Big) ,
\nonumber \\
\hat{n} = n - \frac{1}{2} \theta^{\rho \sigma} A_\rho \partial_\sigma n\,,
\;\;\;\;\;\
\hat{\alpha} = \alpha - \frac{1}{2} \theta^{\rho \sigma} A_\rho \partial_\sigma \alpha
\label{SWM}
\end{eqnarray}

this map transforms a noncommutative gauge orbit in a commutative one. Following the reference \cite{ghosh} we substitute the above form of $\hat{n}$ and $\hat{A_\mu}$ in the action (\ref{NCP1}), which reads

\begin{eqnarray}
\hat{S} &= \int d^3 x  \Big[|D_\mu n|^2
+ \frac{1}{2} \theta^{\alpha \beta } \lbrace F_{\alpha \mu}\left( (D_\beta n)^\dagger D^\mu n +  (D^\mu n)^\dagger D_\beta n \right) \nonumber \\ &
- \frac{1}{2} F_{\alpha \beta} (D^\mu n)^\dagger D_\mu n \rbrace \Big]
\label{SW}
\end{eqnarray}

As in reference \cite{ghosh} we assume

\begin{eqnarray}
\hat{n}^\dagger \hat{n} = n^\dagger n =1
\end{eqnarray}

The equation of motion, derived from the action (\ref{SW}), for the gauge field is

\begin{eqnarray}
2i(-iA_\mu  + n^\dagger \partial_\mu n) \Big(1- \frac{1}{2} \theta^{\alpha \beta} F_{\alpha \beta}\Big) \nonumber \\
+ \frac{1}{2} \theta_{\alpha \mu} \Big[ \partial^\alpha \lbrace (D^\beta n)^\dagger D_\beta n\rbrace - \partial_\beta \lbrace (D^\alpha n)^\dagger D^\beta n + (D^\beta n)^\dagger D^\alpha n \rbrace \Big] \nonumber \\
- \frac{1}{2} \theta_{\alpha \beta} \Big[ \partial^\alpha \lbrace (D^\beta n)^\dagger D_\mu n + (D_\mu n)^\dagger  D^\beta n + i F_\mu^\alpha (n^\dagger D^\beta n - (D^\beta n)^\dagger n)\rbrace \Big] =0
\label{ESW}
\end{eqnarray}
This equation can be rewritten in a reduced form as

\begin{eqnarray}
A_\mu =  -i n^\dagger \partial_\mu n + a_\mu (\theta)
\label{ER}
\end{eqnarray}
where $a_\mu (\theta)$ represent the $O(\theta)$ correction obtained from (\ref{ESW}). Nevertheless, as it is mentioned in the paper of Ghosh  \cite{ghosh}, the term $a_\mu (\theta)$ does not play any role when we introduce (\ref{ER}) in (\ref{SW}) because, here, we are only interested on the first order correction in $\theta$. Thus we drop the term $a_\mu (\theta)$ in (\ref{ER}) and identify

\begin{eqnarray}
A_\mu =  -i n^\dagger \partial_\mu n \,,
\;\;\;\;\;\ n^\dagger n =1
\label{}
\end{eqnarray}

Here we are interested in time-independent instanton solutions to the field equations that ensure the finiteness of the action (\ref{SW}). These are stationary points of the energy, which for static field configuration reads as

\begin{eqnarray}
E = -\hat{S} =& \int d^2 x  \Big[|D_i n|^2 - A_0^2
- \frac{1}{2} \theta^{1 2} \lbrace \partial_1 A_0^2 (i n^\dagger \partial_2 n + A_2) - \partial_2 A_0^2 (i n^\dagger \partial_1 n + A_1) \nonumber \\ &
- F_{1 2} \left(|D_i n|^2 + A_0^2 \right) \rbrace \Big]
\label{SWS}
\end{eqnarray}

By varying with respect to $A_0$ we obtain the relation

\begin{eqnarray}
A_0 =   a_0 (\theta)
\label{ERS}
\end{eqnarray}
which, in virtue of our considerations, may be rewritten as $A_0 =  0$. Replacing this relation in the action (\ref{SWS})  we arrive to

\begin{eqnarray}
E =& \int \,\,d^2 x  |D_i n|^2 \Big( 1
+ \frac{1}{2} \theta^{1 2} F_{1 2} \Big)
\label{31}
\end{eqnarray}

Since we are not interested in $O(\theta^2)$ contribution, the last expression may be written as

\begin{eqnarray}
E =& \int \,\,d^2 x  |D_i n|^2 \Big( 1
+ \frac{1}{4} \theta^{1 2} F_{1 2} \Big)^2
\label{}
\end{eqnarray}
which show that the energy is zero or a positive number. The finiteness of the energy requires that the covariant derivative must vanish asymptotically. This fixes the asymptotic behavior of the fields

\begin{eqnarray}
\lim_{r \to \infty} n(x) = n^0 e^{i\alpha(\phi)}
\,,
\;\;\;\;\;\
\lim_{r \to \infty} A_i = \partial_i \alpha
\end{eqnarray}

where $n^0$ is a fixed complex vector with $(n^0)^\dagger n^0 = 1$ and $\alpha$ is common phase angle. This $\alpha$ depend on  $\phi$, the angle in coordinate space that parameterizes the boundary of the space. With these conditions the magnetic flux reads

\begin{eqnarray}
\Phi =& \int \,\,d^2 x B = \oint_{|x|=\infty} \,\, A_i dx^i  = 2\pi N
\label{}
\end{eqnarray}
where $N$ is a topological invariant which takes only integer values.

As in the commutative $CP(1)$ model we can use to explore the minimum of the energy the following identity

\begin{eqnarray}
|D_i n|^2 =   |( D_1 \pm iD_2)n|^2 \pm B
\end{eqnarray}

Using this identity the energy functional becomes

\begin{eqnarray}
E =& \int \,\,d^2 x \Big[ \Big( 1
+ \frac{1}{4} \theta^{1 2} F_{1 2} \Big)^2 |( D_1 \pm iD_2)n|^2 \pm \frac{1}{2} \theta^{1 2}B^2 \Big] \pm B + O(\theta^2)
\label{E}
\end{eqnarray}

Here we choose the upper signs for positive flux and for negative flux we choose the lower signs. Choosing the upper sign we obtain the following lower bound on the energy

\begin{eqnarray}
E \geq \Phi + \int \,\,d^2 x \frac{1}{2} \theta^{1 2}B^2
\label{Emas}
\end{eqnarray}
which is saturated by fields satisfying the first-order Bogomol'nyi self-duality equation \cite{bogo}

\begin{eqnarray}
|( D_1 + iD_2)n|^2 =0
\label{cond}
\end{eqnarray}

This is the result arrived in Ref.\cite{ghosh}. However the term  $\Phi + \int \,\,d^2 x \frac{1}{2} \theta^{1 2}B^2$ is not really a lower bound on the energy. We will show that the integral in the equation (\ref{Emas}) goes to zero when the region of integration becomes infinity. We begin the proof by the considering the following ansatz for the $N$-instanton solution

\begin{eqnarray}
n(\phi, r)=  \left( \begin{array}{c}
\cos(\frac{f(r)}{2})e^{i N \phi}\\
\sin(\frac{f(r)}{2} )\end{array} \right)
\,,
\;\;\;\;\;\
A_r =0\;, \nonumber \\
A_\phi (r)= -in^\dagger \partial_\phi n = \frac{N}{r}cos^2(\frac{f(r)}{2})
\label{anz}
\end{eqnarray}

where $r=\sqrt{x_1^2 + x_2^2}$ and $\phi = tan^{-1} (\frac{x_2}{x_1})$ . Introducing this ansatz in the equation (\ref{E}) and taken the upper sign,  we obtain

\begin{eqnarray}
E (R)= 2\pi \int_0^R \,\,d r\,\, r \Big[ \Big(
1- \frac{1}{4} \theta^{1 2} N \frac{\sin(f(r))\partial_r f(r)}{2r}  \Big)^2 \Big( \frac{1}{4}(\partial_r f(r))^2 \nonumber \\
+ N \frac{\sin(f(r))\partial_r f(r)}{2r}
 +\Big(N\frac{\sin(f(r))}{2r}\Big)^2 \Big)  + \frac{1}{2} \theta^{1 2} \Big(N \frac{\sin(f(r))\partial_r f(r)}{2r}\Big)^2\Big]  + \Phi,
\label{E2}
\end{eqnarray}
whereas the substitution in the equation (\ref{cond}) gives

\begin{eqnarray}
\frac{1}{4}(\partial_r f(r))^2
+ N \frac{\sin(f(r))\partial_r f(r)}{2r}
 +\Big(N\frac{\sin(f(r))}{2r}\Big)^2 =0
\label{41}
\end{eqnarray}

In (\ref{E2}) we have integrated over a two-dimensional disc $D_R$ of radius $R$. Also we have considered only the corrections to first order in  $\theta$.
From here it follows that the appropriate boundary conditions for finite energy solution are\cite{my}

\begin{eqnarray}
\lim_{r \to 0} f(r) = \pi
\,,
\;\;\;\;\;\
\lim_{r \to R} f(r) = 0
\label{bound}
\end{eqnarray}

Following the reference \cite{my} we shall denote  the solution of Eq.(\ref{41}) subject to the boundary conditions  (\ref{bound}) on a disc $D_{R_1}$ as $f_{R_1} (r)$. Consider now the following configuration defined on a disc $D_{\lambda R_1}$

\begin{eqnarray}
\tilde{f}_{\lambda R_1}(r) = f_{R_1}(\frac{r}{\lambda})
\label{conf}
\end{eqnarray}

where $\lambda > 1$. According to (\ref{bound}) this configuration is subject to the following boundary conditions

\begin{eqnarray}
\lim_{r \to 0} \tilde{f}_{\lambda R_1}(r) = f_{R_1}(0) = \pi
\,,
\;\;\;\;\;\
\lim_{r \to \lambda R_1} \tilde{f}_{\lambda R_1}(r) = f_{R_1}(R_1) = 0
\label{bound1}
\end{eqnarray}

We can evaluate the energy (\ref{E2}) for the configuration (\ref{conf}) on the disc $D_{\lambda R_1}$

\begin{eqnarray}
\tilde{E}(\lambda R_1) = 2\pi \int_0^{\lambda R_1} \,\,d r\,\, r \Big[ \Big(
1- \frac{1}{4} \theta^{1 2} N \frac{\sin(\tilde{f}_{\lambda R_1}(r) )\partial_r \tilde{f}_{\lambda R_1}(r)}{2r}  \Big)^2 \Big( \frac{1}{4}(\partial_r \tilde{f}_{\lambda R_1}(r) )^2 \nonumber \\
+ N \frac{\sin(\tilde{f}_{\lambda R_1}(r) )\partial_r \tilde{f}_{\lambda R_1}(r) }{2r}
 +\Big(N\frac{\sin(\tilde{f}_{\lambda R_1}(r) )}{2r}\Big)^2 \Big)  + \frac{1}{2} \theta^{1 2} \Big(\frac{\sin(\tilde{f}_{\lambda R_1}(r) )\partial_r \tilde{f}_{\lambda R_1}(r) }{2r}\Big)^2\Big]  + \Phi
\label{E3}
\end{eqnarray}

Since $\tilde{f}_{\lambda R_1}(r)$ satisfies the same boundary conditions as the solution of Eq.(\ref{41}) on the disc $D_{\lambda R_1}$, which we denote here as $f_{\lambda R_1}(r)$, we arrive to

\begin{eqnarray}
\Phi \leq E(\lambda R_1) \leq \tilde{E}(\lambda R_1)
\label{inq}
\end{eqnarray}

where $E(\lambda R_1)$ denote the functional energy evaluated on the disc $D_{\lambda R_1}$ for the solution $f_{\lambda R_1}(r)$. Under the transformation $r=x\lambda$ the functional  (\ref{E3}) becomes

\begin{eqnarray}
\tilde{E}(\lambda R_1) = 2\pi \int_0^{R_1} \,\,d x \,\,x \Big[ \Big(
1- \frac{1}{4\lambda^2} \theta^{1 2} N \frac{\sin(f_{R_1}(x) )\partial_x f_{ R_1}(x)}{2x}  \Big)^2 \Big( \frac{1}{4}(\partial_x f_{ R_1}(x) )^2 \nonumber \\
+ N \frac{\sin(f_{R_1}(x) )\partial_x f_{ R_1}(x) }{2x} + \Big(N\frac{\sin(f_{ R_1}(x) )}{2x}\Big)^2
\Big)  + \frac{1}{2\lambda^2} \theta^{1 2} \Big(\frac{\sin(f_{ R_1}(x) )\partial_x f_{ R_1}(x) }{2x}\Big)^2\Big]  + \Phi
\label{}
\end{eqnarray}

Since $f_{R_1}$ is a solution of (\ref{41}) on the disc $D_{R_1}$, the functional $\tilde{E}(\lambda R_1)$  reduce to

\begin{eqnarray}
\tilde{E}(\lambda R_1) = 2\pi \int_0^{R_1} \,\,d x \,\, x\Big[   \frac{1}{2\lambda^2} \theta^{1 2} (\frac{\sin(f_{ R_1}(x) )\partial_x f_{ R_1}(x) }{2x})^2\Big]  + \Phi
\label{}
\end{eqnarray}

The limit for arbitrary large $\lambda$ gives

\begin{eqnarray}
\lim_{\lambda \to \infty} \tilde{E}(\lambda R_1)= \Phi
\end{eqnarray}

which implies, in virtue of equation (\ref{inq}), that

\begin{eqnarray}
\lim_{\lambda \to \infty} E(\lambda R_1)= \Phi
\end{eqnarray}

Thus the topological bound is saturated for arbitrary large $R$ and for the fields satisfying the condition (\ref{cond}).
This also shows that both the magnetic field and the integral of its square should be zero for arbitrary large $R$. Nevertheless the magnetic flux remains constant for all two-dimensional disc. This is because the magnetic flux depends only on the boundary conditions which are fixed

\begin{eqnarray}
\int B dx^2 = 2\pi \int_0^R dr r B= 2\pi \int_0^R dr r \frac{\partial_r (r A_\phi)}{r} =\left.r A_\phi \right|_0^R =N\cos^2(\frac{f(R)}{2})=2\pi N
\label{51}
\end{eqnarray}

As $R\rightarrow \infty$, $B\rightarrow 0$ such that

\begin{eqnarray}
\lim_{R \to \infty} \int^R_0 B (r, R) r dr= 2\pi N
\end{eqnarray}

One important aspect here is that the solution of our model, as $R\rightarrow \infty$, is the same as the infinite size solution found in Ref.\cite{my} for the commutative Chern-Simons-$CP(1)$ system. Both systems are subject to identical boundary conditions and satisfied the same BPS equations. Thus the effect of mapping to a first order in $\theta$ the noncommutative $CP(1)$ theory by Seiberg-Witten map is the same as to add a Chern-Simons term in the commutative $CP(1)$ theory.

By varying the action (\ref{SW}) with respect to $CP(1)$ field $n$ yields

\begin{eqnarray}
D^\mu \Big[ \Big(1- \frac{1}{4} \theta^{\alpha \beta} F^{\alpha \beta}\Big) D_\mu n \Big]\nonumber \\
+ \frac{1}{2} \theta_{\alpha \mu} \Big[ i D_\alpha \lbrace (D^\sigma n)^\dagger D_\sigma n D_\beta n\rbrace + D_\beta \lbrace F_{\alpha \mu} D^\mu n \rbrace + D^\mu \lbrace F{\alpha \mu} D_{\beta} n\rbrace \nonumber \\
-iD_\alpha \lbrace \Big( (D_\beta n)^\dagger D^\mu n + (D^\mu n)^\dagger D_\beta n \Big)D_\mu n \rbrace \nonumber \\
+iD_\mu \lbrace \Big( (D_\beta n)^\dagger D^\mu n + (D^\mu n)^\dagger D_\beta n \Big)D_\alpha n \rbrace \Big] -\lambda n =0
\label{D_1}
\end{eqnarray}

Following the reference \cite{ghosh} we may eliminate the lagrange multiplier $\lambda$

\begin{eqnarray}
\lambda = n^\dagger D^i D_i  n + \frac{1}{2} \theta^{1 2} n^\dagger (F_{1 2} D_i n)
\end{eqnarray}
Inserting $\lambda$ in the equation of motion we get

\begin{eqnarray}
 D^i D_i  n - \Big( n^\dagger D^i D_i n \Big) n + \frac{1}{2} \theta^{1 2}\Big[D^i( F_{1 2} D_i n) - n^\dagger D^i (F_{1 2} D_i n) n \Big] = 0
 \label{jji}
\end{eqnarray}

As we mentioned the magnetic field $F_{1 2}$ becomes zero as $R\rightarrow \infty$. So the equation (\ref{jji}) reduce to

\begin{eqnarray}
D^i D_i  n - \Big( n^\dagger D^i D_i n \Big) n = 0
\label{cond2}
\end{eqnarray}

This is the Euler-Lagrange equation of the $CP(1)$ theory \cite{witten}. Hence the BPS and the Euler-Lagrange equations are identical to those of the commutative $CP(1)$ system and there are no incompatibility between them. It is interesting to note that our result is valid only when we are working in two-dimensional infinite disc. For finite space the BPS and the field equations are no compatible and the result obtained by Ghosh is valid.

\section{ Noncommutative Chern-Simons-$CP(1)$ case}

In this section we analyze the  model obtained by the application of the Seiberg-Witten map to the noncommutative Chern-Simons-$CP(1)$ action. We show that this model admit the same instanton solution as the system studied in the previous section.  The action that describe this model is

\begin{eqnarray}
\hat{S}= \int d^3 x \Big[ \kappa \epsilon^{\mu \nu \lambda} \Big( \hat{A}_\mu \partial_\nu \hat{A}_\lambda + \frac{2i}{3} \hat{A}_\mu \hat{A}_\nu \hat{A}_\lambda \Big) + |\hat{D}_\mu \hat{n}|^2 \Big]
\end{eqnarray}

Under the Seiberg-Witten map (\ref{SWM}), this action transforms as follows

\begin{eqnarray}
\hat{S}= \int d^3 x \Big[\kappa \epsilon^{\mu \nu \lambda} A_\mu \partial_\nu A_\lambda + |D_\mu n|^2 + \frac{1}{2} \theta^{\alpha \beta } \lbrace F_{\alpha \mu}\left( (D_\beta n)^\dagger D^\mu n +  (D^\mu n)^\dagger D_\beta n \right) \nonumber \\
- \frac{1}{2} F_{\alpha \beta} (D^\mu n)^\dagger D_\mu n \rbrace \Big]
\label{SCS}
\end{eqnarray}

This action differs from the action (\ref{SW}) only on the commutative Chern-Simons term \cite{GS}. Using the notation of equation  (\ref{ER}) we can write the equation of motion for the gauge field as

\begin{eqnarray}
A_\mu = -i n^\dagger \partial_\mu n -\frac{\kappa}{2} \epsilon_{\mu \nu \rho } F^{\nu \rho} + a_\mu (\theta)
\label{A_1}
\end{eqnarray}
whereas the the dynamical equation for the $CP(1)$ field is the same as the equation obtained in the previous section
\begin{eqnarray}
D^\mu \Big[ \Big(1- \frac{1}{4} \theta^{\alpha \beta} F^{\alpha \beta}\Big) D_\mu n \Big]\nonumber \\
+ \frac{1}{2} \theta_{\alpha \mu} \Big[ i D_\alpha \lbrace (D^\sigma n)^\dagger D_\sigma n D_\beta n\rbrace + D_\beta \lbrace F_{\alpha \mu} D^\mu n \rbrace + D^\mu \lbrace F{\alpha \mu} D_{\beta} n\rbrace \nonumber \\
-iD_\alpha \lbrace \Big( (D_\beta n)^\dagger D^\mu n + (D^\mu n)^\dagger D_\beta n \Big)D_\mu n \rbrace \nonumber \\
+iD_\mu \lbrace \Big( (D_\beta n)^\dagger D^\mu n + (D^\mu n)^\dagger D_\beta n \Big)D_\alpha n \rbrace \Big] -\lambda n =0
\label{D_11}
\end{eqnarray}

In the static field configuration the action (\ref{SCS}) leads to the following form of the functional energy

\begin{eqnarray}
E = \int d^2 x \Big[-2\kappa A_0 F_{1 2} + |D_i n|^2 - A_0^2 +
\frac{1}{2} \theta^{1 2} \lbrace \partial_1 A_0^2 (i n^\dagger \partial_2 n + A_2 ) - \nonumber \\
\partial_2 A_0^2 (i n^\dagger \partial_1 n + A_1 ) - F_{1 2}\left( |D_i n|^2 + A_0^2 \right)
\rbrace \Big]
\label{}
\end{eqnarray}

The substitution of the time component of  (\ref{A_1}) in the expression of the energy yields

\begin{eqnarray}
E = \int d^2 x \Big[\kappa F^2_{1 2} + |D_i n|^2 -
\frac{1}{2} \theta^{1 2} \lbrace \kappa^2 \partial_1 F^2_{1 2} (i n^\dagger \partial_2 n + A_2 ) - \nonumber \\
\kappa^2 \partial_2 F^2_{1 2} (i n^\dagger \partial_1 n + A_1 ) - F_{1 2}\left( |D_i n|^2 + \kappa^2 F^2_{1 2} \right)
\rbrace \Big]
\label{}
\end{eqnarray}

After integration by parts, the last equation can be rewritten as

\begin{eqnarray}
E = \int d^2 x \Big[\kappa F^2_{1 2} + |D_i n|^2 -
\frac{1}{2} \theta^{1 2} \lbrace -\kappa^2  F^2_{1 2} \epsilon^{i j}\Big(i\partial_i (n^\dagger \partial_j n) + \partial_i A_j \Big) - \nonumber \\
F_{1 2}\left( |D_i n|^2 + \kappa^2 F^2_{1 2} \right)
\rbrace \Big]
\label{}
\end{eqnarray}
\begin{eqnarray}
E = \int d^2 x \Big[\kappa^2 F^2_{1 2} \Big( 1 +
\frac{1}{2} \theta^{1 2} \lbrace -2\pi J_0^{CP(1)} + 2 F_{1 2}\rbrace \Big) +
|D_i n|^2 \left(1 + \frac{1}{2} \theta^{1 2} F_{1 2} \right)\Big]
\label{ECS}
\end{eqnarray}

Here $J_0^{CP(1)} = -\frac{i}{2\pi} \epsilon^{ij} (D_i n)^\dagger (D^j n)$ is the topological charge density of $CP(1)$ model. For $ \kappa =0$ the last expression reduce to the functional energy (\ref{31}) of the mapped pure $CP(1)$ model. The energy (\ref{ECS}) is further rewritten by using the identity

\begin{eqnarray}
|D_i n|^2 =   |( D_1 \pm iD_2)n|^2 \pm 2\pi J_0^{CP(1)},
\end{eqnarray}
\begin{eqnarray}
E = \int d^2 x \Big[\kappa^2 F^2_{1 2} \Big( 1 +
\frac{1}{4} \theta^{1 2} \lbrace -2\pi J_0^{CP(1)} + 2 F_{1 2}\rbrace \Big)^2 + \nonumber \\
|( D_1 \pm iD_2)n|^2 \left(1 + \frac{1}{4} \theta^{1 2} F_{1 2} \right)^2 \pm \pi \theta^{1 2}  J_0^{CP(1)} F_{1 2}\Big] \pm 2\pi \int d^2 x J_0^{CP(1)} + O(\theta^2)
\label{64}
\end{eqnarray}
where $2\pi \int d^2 x J_0^{CP(1)}$ is connected with magnetic flux by following formula \cite{my}
\begin{eqnarray}
2\pi \int d^2 x J_0^{CP(1)} = \Phi - \frac{1}{2}\oint J_i dx^i
\end{eqnarray}
The line integral vanishes for any finite solution on the plane.
We will choose the upper sings for positive flux and the lower signs for negative flux. We shall show that the magnetic flux is lower bound on the energy as the region of integration becomes infinity. This bound will be saturated by static configuration obeying the self-duality equation

\begin{eqnarray}
|( D_1 \pm iD_2)n|^2 =0
\label{ojo}
\end{eqnarray}

Let us consider the following ansatz for the $N$-instanton solution \cite{my}

\begin{eqnarray}
n(\phi, r)=  \left( \begin{array}{c}
\cos(\frac{f(r)}{2})e^{i N \phi}\\
\sin(\frac{f(r)}{2} )\end{array} \right)
\,,
\;\;\;\;\;\
 A_\phi (r)= a(r)
\,,
\;\;\;\;\;\
A_r =0
\label{ansatz}
\end{eqnarray}

where $r$ and $\phi$ are defined as in the Section $3$. Using this ansatz it is not very difficult to show that the $N$-instanton energy functional (\ref{64}) reads as

\begin{eqnarray}
E(R) = 2\pi\int^R_0 dr\,\, r \Big[\kappa^2 \Big(\frac{\partial_r (r a(r))}{r}\Big)^2 \Big( 1 +
\frac{1}{4} \theta^{1 2} \lbrace \frac{N}{2r} \sin(f(r)) \partial_r f(r) +\nonumber \\
 2\frac{\partial_r (r a(r))}{r} \rbrace \Big)^2 +
\Big(\frac{(\partial_r f(r))^2}{4} \pm \frac{\sin (f(r))}{2r} \partial_r f(r) +\cos^2 (\frac{f(r)}{2})\frac{N}{r}\Big(\frac{N}{r} -2 a(r)  \Big) \nonumber \\
+ a^2 (r)\Big)
\left(1 + \frac{1}{4} \theta^{1 2} \frac{\partial_r (r a(r))}{r} \right)^2
\mp  \theta^{1 2} \frac{N}{4}  \frac{\partial_r (r a(r))}{r^2}  \sin(f(r)) \partial_r f(r)\Big] + |\Phi|
\label{ECS1}
\end{eqnarray}
where we integrate over a two-dimensional disc $D_R$. The equation (\ref{ojo}) takes the form
\begin{eqnarray}
\frac{(\partial_r f(r))^2}{4} \pm \frac{\sin (f(r))}{2r} \partial_r f(r) +\cos^2 (\frac{f(r)}{2})\frac{N}{r}\Big(\frac{N}{r} - 2a(r) \Big) + a^2 (r) =0
\label{eojo}
\end{eqnarray}

It is reasonable to impose for finiteness  of the energy the following boundary conditions \cite{my}

\begin{eqnarray}
\lim_{r \to 0} f(r) = \pi
\,,
\;\;\;\;\;\
\lim_{r \to 0} a(r) = 0
\label{b1}
\end{eqnarray}

\begin{eqnarray}
\lim_{r \to R} f(r) = 0
\,,
\;\;\;\;\;\
\lim_{r \to R}a(r) =\frac{ N}{R}
\label{b2}
\end{eqnarray}

Notice that the boundary condition for $f(r)$ are the same as the conditions (\ref{bound}), whereas the $A_\phi (r)$ defined in the ansatz (\ref{anz}) satisfies the boundary conditions (\ref{b1}) and (\ref{b2}) imposed to $a(r)$.
As in Ref.\cite{my} we consider $a_{R_1}(r)$ and $f_{R_1}(r)$ the solutions of (\ref{eojo}) on a disc $D_{R_1}$, subject to the boundary conditions (\ref{b1}) and (\ref{b2}). With these solutions we construct the following configuration defined on a disc $D_{\lambda R_1}$ \cite{my}

\begin{eqnarray}
\tilde{a}_{\lambda R_1}(r) =\frac{a_{ R_1}(\frac{r}{\lambda})}{\lambda}
\,,
\;\;\;\;\;\
\tilde{f}_{\lambda R_1}(r) = f_{ R_1}(\frac{r}{\lambda})
\label{configur}
\end{eqnarray}

again $\lambda >1$. The boundary conditions for this configuration are

\begin{eqnarray}
\lim_{r \to 0} \tilde{f}_{\lambda R_1}(r) = \pi
\,,
\;\;\;\;\;\
\lim_{r \to 0} \tilde{a}_{\lambda R_1}(r) = 0
\label{bo1}
\end{eqnarray}

\begin{eqnarray}
\lim_{r \to \lambda R_1} \tilde{f}_{\lambda R_1}(r) = 0
\,,
\;\;\;\;\;\
\lim_{r \to \lambda R_1}\tilde{a}_{\lambda R_1}(r) = \frac{a_{R_1}(R_1)}{\lambda}=\frac{ N}{\lambda R_1}
\label{bo2}
\end{eqnarray}

The energy functional (\ref{ECS1}) evaluated in the configuration (\ref{configur}) gives

\begin{eqnarray}
\tilde{E}(\lambda R_1) = 2\pi\int^{\lambda R_1}_0 dr \,\,r \Big[\kappa^2 \Big(\frac{\partial_r (r\tilde{a}_{\lambda R_1}(r) )}{r}\Big)^2 \Big( 1 +
\frac{1}{4} \theta^{1 2} \lbrace \frac{N}{2r} \sin(\tilde{f}_{\lambda R_1}(r)) \partial_r \tilde{f}_{\lambda R_1}(r) +\nonumber \\
 2\frac{\partial_r (r \tilde{a}_{\lambda R_1}(r))}{r} \rbrace \Big)^2 +
\Big(\frac{(\partial_r \tilde{f}_{\lambda R_1}(r))^2}{4} \pm \frac{\sin (\tilde{f}_{\lambda R_1}(r))}{2r} \partial_r \tilde{f}_{\lambda R_1}(r) +\nonumber \\
\cos^2 (\frac{\tilde{f}_{\lambda R_1}(r)}{2})\frac{N}{r}\Big(\frac{N}{r} - 2\tilde{a}_{\lambda R_1}(r)  \Big) + \tilde{a}^2_{\lambda R_1}(r)\Big)
\left(1 + \frac{1}{4} \theta^{1 2} \frac{\partial_r (r \tilde{a}_{\lambda R_1}(r))}{r} \right)^2  \nonumber \\
\mp  \theta^{1 2} \frac{N}{4}  \frac{\partial_r (r \tilde{a}_{\lambda R_1}(r))}{r^2} \sin(\tilde{f}_{\lambda R_1}(r)) \partial_r \tilde{f}_{\lambda R_1}(r)\Big] + |\Phi|
\label{ECS2}
\end{eqnarray}
We will denote the solution of (\ref{eojo}) corresponding to the disc $D_{\lambda R_1}$ and subject to the boundary conditions (\ref{b1})-(\ref{b2}) as $a_{\lambda R_1}(r)$ and $f_{\lambda R_1}(r)$ and its energy as $E(\lambda R_1)$. Then it is clear that the configuration (\ref{configur}) do not minimize the energy on the disc $D_{\lambda R_1}$.
From this and the equation (\ref{64}) we can write

\begin{eqnarray}
|\Phi| \pm  \pi \theta^{1 2} \int d^2 x J_0^{CP(1)} F_{1 2} \leq E(\lambda R_1) \leq \tilde{E}(\lambda R_1)
\label{inq2}
\end{eqnarray}

Again we introduce the transformation $r= x\lambda$

\begin{eqnarray}
\tilde{E}(\lambda R_1) = 2\pi\int^{R_1}_0 dx \,\,x \Big[\frac{\kappa^2}{\lambda^2} \Big(\frac{\partial_x (x a_{R_1})}{x}\Big)^2 \Big( 1 +
\frac{1}{4\lambda^2} \theta^{1 2} \lbrace \frac{N}{2x} \sin(f_{R_1}) \partial_x f_{ R_1} +\nonumber \\
 2\frac{\partial_x (a_{R_1})}{x} \rbrace \Big)^2 +
\Big(\frac{(\partial_x f_{ R_1})^2}{4} \pm \frac{\sin (f_{R_1})}{2x} \partial_x f_{R_1} +\nonumber \\
\cos^2 (\frac{f_{R_1}}{2})\frac{N}{x}\Big(\frac{N}{x} - 2a_{R_1} \Big) + a^2_{R_1}\Big)
\left(1 + \frac{1}{4\lambda^2} \theta^{1 2} \frac{\partial_x (x a_{R_1})}{x} \right)^2  \nonumber \\
\mp  \theta^{1 2} \frac{N}{4\lambda^2}  \frac{\partial_x (x a_{R_1})}{x^2}  \sin(f_{R_1}) \partial_x f_{R_1} \Big] + |\Phi|
\label{ECS3}
\end{eqnarray}

The limit for large $\lambda$ gives

\begin{eqnarray}
\lim_{\lambda \to \infty} \tilde{E}(\lambda R_1)= 2\pi\int^{R_1}_0 dx \,\,x \Big[\frac{(\partial_x f_{ R_1})^2}{4} \pm \frac{\sin (f_{R_1})}{2x} \partial_x f_{R_1} +\nonumber \\
\cos^2 (\frac{f_{R_1}}{2})\frac{N}{x}\Big(\frac{N}{x} -2 a_{R_1}  \Big) + a^2_{R_1}\Big] + |\Phi|
\end{eqnarray}

Since $f_{R_1}$ is a solution of (\ref{eojo}) on the disc $D_{R_1}$ the last expression reduce to

\begin{eqnarray}
\lim_{\lambda \to \infty} \tilde{E}(\lambda R_1)=  |\Phi|
\label{lim}
\end{eqnarray}

In order to complete the proof we show that $\int d^2 x J_0^{CP(1)} F_{1 2}$ becomes zero as $\lambda\rightarrow \infty$. In terms of the solution on the disc $D_{\lambda R_1}$ this integral reads

\begin{eqnarray}
\int d^2 x J_0^{CP(1)} F_{1 2} =  \int_0^{\lambda R_1} dr \,\,r N  \frac{\partial_r (r a_{\lambda R_1}(r))}{2 r^2}  \sin(f_{\lambda R_1}(r)) \partial_r f_{\lambda R_1}(r)
\end{eqnarray}

Introducing the change of variable $r= \lambda x$, we obtain

\begin{eqnarray}
\int d^2 x J_0^{CP(1)} F_{1 2} = \frac{N}{\lambda} \int_0^{ R_1} dx \,\,x   \frac{\partial_x (x a_{\lambda R_1})}{2 x^2} \sin(f_{\lambda R_1}) \partial_x f_{\lambda R_1}
\label{J}
\end{eqnarray}
In a shorten form this integral may be rewritten as

\begin{eqnarray}
\int d^2 x J_0^{CP(1)} F_{1 2} = \frac{N}{\lambda}K(R_1)
\end{eqnarray}

where $K(R_1)$ is finite since

\begin{eqnarray}
K(R_1) \leq E(\lambda R_1)
\end{eqnarray}

Therefore for arbitrary large $\lambda$ the equation (\ref{J}) reads

\begin{eqnarray}
\lim_{\lambda \to \infty} \int d^2 x J_0^{CP(1)} F_{1 2} =  0
\end{eqnarray}

Finally this implies in virtue of equations (\ref{lim}) and (\ref{inq2}) that

\begin{eqnarray}
\lim_{\lambda \to \infty} E(\lambda R_1) =  |\Phi|
\end{eqnarray}

In the same form as the model studied in Section $3$, here the saturation of topological bound requires that the field satisfy the self-duality equation (\ref{ojo}) as the region of integration in (\ref{ECS1}) becomes infinite. In addition, notice that since the boundary conditions are the same in both models, the solution of the model analyzed in Section $3$ is also the $N$-instanton solution for this model.

\section{ Conclusions}

In this article we have discussed a model obtained by the application of the Seiberg-Witten map to noncommutative $CP(1)$ system. This model was studied previously by Ghosh in reference \cite{ghosh}. The author showed there that the BPS equations for this model are the same as those found in the commutative $CP(1)$ model. Nevertheless he claims that the variational equations in both models are different. The difference consists in the $\theta$-term $\frac{1}{2} \theta^{1 2} \partial^i F_{1 2} D_i n$. In the present paper we showed that this term vanishes if boundary conditions are in an infinite plane. More specifically we showed that

\begin{eqnarray}
\lim_{R \to \infty} B(r, R) = 0
\,,
\;\;\;\;\;\
\lim_{R \to \infty}\int^R_0 B^2 (r, R) r dr= 0
\label{}
\end{eqnarray}

This allows us to correct the result present in \cite{ghosh} and also show that for the fields satisfying first order self-duality equations

\begin{eqnarray}
|( D_1 \pm iD_2)n|^2 =0
\label{}
\end{eqnarray}

the magnetic flux is the lower bound on the energy

\begin{eqnarray}
\lim_{R \to \infty} E(R) = \Phi
\end{eqnarray}

It is also remarkable that this model admits, as  $R\rightarrow \infty$, the same infinite size $N$-instanton solutions found in \cite{my}.
Finally, using the method of Section $3$, we have studied the noncommutative  Chern-Simons-$CP(1)$ model, finding that this model also presents the infinite size instanton solutions of the commutative  Chern-Simons-$CP(1)$ system.

\vspace{2cm}
{\bf Acknowledgements}
This work was partially supported by CONICET.

\end{document}